\documentclass{article}

\usepackage{PRIMEarxiv}

\usepackage[acronym,nomain,nonumberlist]{glossaries}
\makeglossaries

\newacronym{jgat}{JGAT}{Joint kernel Graph Attention Network}
\newacronym{fc}{FC}{Functional Connectivity}
\newacronym{sc}{SC}{Structural Connectivity}
\newacronym{gat}{GAT}{Graph Attention Network}
\newacronym{gnn}{GNNs}{Graph Neural Networks}
\newacronym{fmri}{fMRI}{functional Magnetic Resonance Images}
\newacronym{dwi}{DWI}{Diffusion Weighted Imaging}
\newacronym{hcp}{HCP}{Human Connectome Project}
\newacronym{as}{AS}{Attention Scores}
\newacronym{fs}{FS}{Frame Scores}
\newacronym{ml}{ML}{Machine Learning}
\newacronym{rois}{ROIs}{Region of Interests}
\newacronym{mksvm}{MK-SVM}{multi-kernel support vector machine}
\newacronym{asd}{ASD}{Autism spectrum disorder}
\newacronym{fle}{FLE}{Frontal Lobe Epilepsy}
\newacronym{tle}{TLE}{Temporal Lobe Epilepsy}
\newacronym{rnn}{RNN}{Recurrent Neural Network}
\newacronym{cnn}{CNN}{Convolutional Neural Networks}
\newacronym{wm}{WM}{Working Memory}
\newacronym{mlp}{MLP}{Multilayer Perceptron}
\newacronym{gcn}{GCN}{Graph Convolutional Networks}
\newacronym{lstm}{LSTM}{Long short-term memory}
\newacronym{stgcn}{ST-GCN}{Spatial Temporal Graph Convolutional Networks}
\newacronym{stndt}{STNDT}{SpatioTemporal Neural Data Transformer}
\newacronym{ndt}{NDT}{Neural Data Transformer}
\newacronym{adbnn}{ADB-NN}{Attention-Diffusion-Bilinear Neural Network}
\newacronym{sem}{SEM}{standard error of mean}
\newacronym{lh}{LH}{Left Hemisphere}
\newacronym{rh}{RH}{Right Hemisphere}
\newacronym{som}{SomMot}{Somatosensory/Motor}
\newacronym{dor}{DorsAttn}{Dorsal Attention}
\newacronym{vent}{SalVentAttn}{Dorsal Attention}
\newacronym{pfc}{PFC}{Prefrontal cortex}

\usepackage[utf8]{inputenc} 
\usepackage[T1]{fontenc}    
\usepackage{hyperref}       
\usepackage{url}            
\usepackage{booktabs}       
\usepackage{amsfonts}       
\usepackage{nicefrac}       
\usepackage{microtype}      
\usepackage{lipsum}
\usepackage{fancyhdr}       
\usepackage{graphicx}       
\usepackage{subcaption}
\usepackage{float}
\graphicspath{{media/}}     

\title{JGAT: a joint spatio-temporal graph attention model for brain decoding
}

\author{
  Han Yi Chiu \\
  Georgia Institute of Technology \\
  Atlanta, GA 30332 \\
  \texttt{r28875148@gmail.com} \\
   \And
    Liang Zhao\\
    Emory University\\
    Atlanta, GA 30322 \\
    \texttt{liang.zhao@emory.edu} \\
  \And
    Anqi Wu \\
    Georgia Institute of Technology\\
    Atlanta, GA 30332 \\
    \texttt{anqiwu@gatech.edu} \\
}

\begin{document}
\maketitle

\begin{abstract}
    The decoding of brain neural networks has been an intriguing topic in neuroscience for a well-rounded understanding of different types of brain disorders and cognitive stimuli. Integrating different types of connectivity, e.g., Functional Connectivity (FC) and Structural Connectivity (SC), from multi-modal imaging techniques can take their complementary information into account and therefore have the potential to get better decoding capability. However, traditional approaches for integrating FC and SC overlook the dynamical variations, which stand a great chance to over-generalize the brain neural network. In this paper, we propose a Joint kernel Graph Attention Network (JGAT), which is a new multi-modal temporal graph attention network framework. It  integrates the data from functional Magnetic Resonance Images (fMRI) and Diffusion Weighted Imaging (DWI) while preserving the dynamic information at the same time. We conduct brain-decoding tasks with our JGAT on four independent datasets: three of 7T fMRI datasets from the Human Connectome Project (HCP) and one from animal neural recordings. Furthermore, with Attention Scores (AS) and Frame Scores (FS) computed and learned from the model, we can locate several informative temporal segments and build meaningful dynamical pathways along the temporal domain for the HCP datasets. The URL to the code  of JGAT model: \url{https://github.com/BRAINML-GT/JGAT}.
\end{abstract}

\keywords{Brain decoding \and Functional Connectivity \and Structural Connectivity \and Machine learning}

\section{Introduction}
Analysis of brain neural networks measured by neuroimaging techniques has helped in revealing potential structures and functions of human brains \cite{cbn}. Specifically, these structures and functions of human brains can provide informative representations and patterns which play a pivotal role in identifying multiple brain neural diseases \cite{brain2, brain3, brain4, brain5, brain6, brain7, brain8} and understanding human behavior and cognition. The functional and structural information measured by noninvasive neuroimaging technologies is often referred to as \gls{fc} and \gls{sc}. Researchers have developed a variety of \gls{ml} models to integrate \gls{fc} and \gls{sc} for decoding the brain networks of brain disorders and cognitive stimuli. \gls{gnn} are the mainstream \gls{ml} 
 models commonly used to integrate \gls{fc} and \gls{sc}. There are existing graph-based models that have been built and dealt with many neurological problems successfully \cite{gmodel1, gmodel2, gmodel3, braingrl, svm, braingnn, adbnn}. Nevertheless, a common issue of these works in neuroimaging is that they only integrate \gls{fc} and \gls{sc} for brain decoding and fail to capture dynamic changes and patterns.

In order to tackle the time information in neuroimaging data, recent literature implements several cutting-edge models for the purpose of both modeling dynamical systems and preserving the spatial relationship \cite{tgcn,stndt,stgcn}. However, these approaches capture temporal embedding and spatial embedding independently, which leads to a limitation in modeling the conditions where a current brain region influences other brain regions in a few previous time steps, i.e., cross-region dynamics. Furthermore, they achieve the spatial-temporal embedding in a sequential way, e.g., deriving the spatial embedding from the input data and then deriving the temporal embedding from the spatial one. This implies that only one of the embeddings relies on the original data inputs, potentially limiting the integration of both spatial and temporal information. We are interested in aggregating input features into both spatial and temporal embeddings jointly. The joint embedding helps build connections between brain regions from different locations and time frames. 

For modeling the aforementioned conditions, we propose a \gls{jgat} to combine \gls{fc} and \gls{sc} for the integration of multi-modal imaging techniques. Besides integrating signals of the statistical dependency across brain regions and their physical locations, we also consider dynamical variations across several time frames. Moreover, by considering temporal and spatial domains at the same time, our approach allows to describe the dynamic cognitive pathways associated with various types of cognitive stimuli. In the evaluation section, we compare our model with both \gls{ml} architectures and some advanced dynamical graph models. We will show that our JGAT model can provide meaningful interpretations for dynamical connectivity and construct pathways of informative segments across the time series of cognitive stimuli for several neuroimaging datasets.

\section{Methodology}
\label{sec:headings}

\subsection{Architecture overview}
\begin{figure}[!t]
    \centering
	\includegraphics[width=1\textwidth]{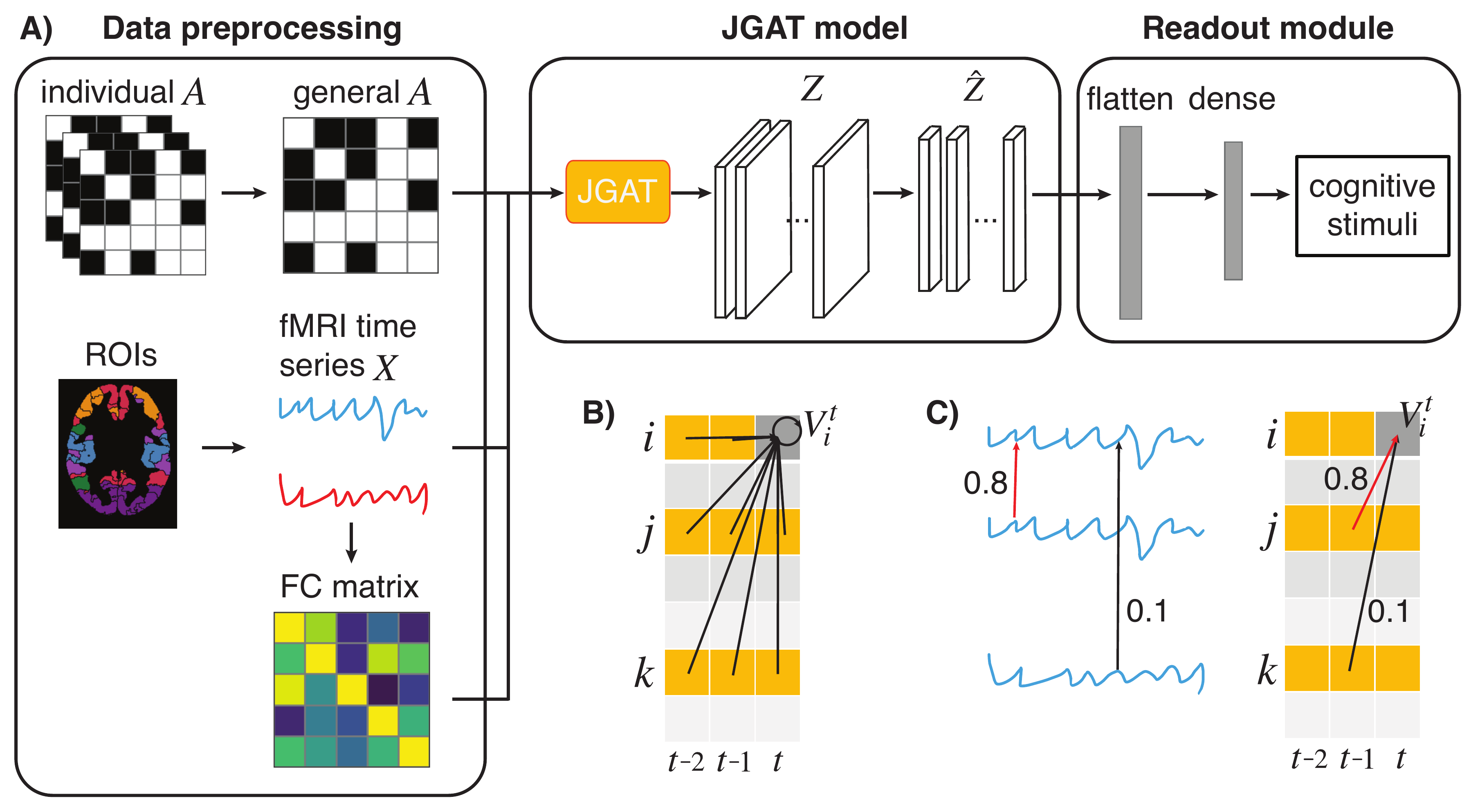}
	\caption{A) Overview of the JGAT architecture. B) Schematic of the joint graph. C) Effect of \gls{fc}.}
\label{fig:jgatmodel}
\end{figure}

Our goal is to perform classification tasks on several cognitive datasets. Therefore, we build a brain decoding architecture (Fig.~\ref{fig:jgatmodel}A) taking neuroimaging data as the input and predicting labels corresponding to the cognitive stimuli. The inputs of the architecture consist of three streams that result from two resources. \gls{dwi} data is the first resource. We can obtain \gls{dwi} data from individual subjects, each providing an SC or adjacency matrix $\tilde{A} \in \mathbb{R}^{N\times N}$. We then calculate a general adjacency $A \in \mathbb{R}^{N\times N}$ from $\tilde{A}$ as the first stream. The second resource is the \gls{fmri} sequence $X \in \mathbb{R}^{N\times T}$, measured from $N$ \gls{rois} at $T$ time frames. It contributes to the other two streams. The second stream is the \gls{fmri} sequence. 
The third stream is the \gls{fc} matrix which is $N$ by $N$, calculated from the \gls{fmri} data by the Pearson correlation coefficient between each pair of brain regions \cite{damoiseaux2006consistent,greicius2003functional,shirer2012decoding,salvador2005neurophysiological}.

Afterward, three streams of input data are fed into the proposed \gls{jgat} model. The \gls{jgat} model consists of a novel \gls{jgat} layer and a dense neural net layer. The \gls{jgat} layer aggregates input features into joint spatial-temporal embedding, denoted as $Z$, via a graph attention mechanism. A dense layer is then applied to $Z$ to reduce the computational complexity and form the downstream embedding $\hat{Z}$. Finally, the readout module for graph classification contains one flatten layer followed by two dense layers, with the $Softmax$ function used in the last dense layer. We also apply two dropout layers in the readout module to prevent overfitting. We demonstrate the overview of \gls{jgat} architecture in Fig.~\ref{fig:jgatmodel}A.

\subsection{Joint Kernel Graph Attention Network}
The core contribution of this paper is the development of the Joint kernel Graph Attention Network (JGAT) that can simultaneously incorporate both temporal and spatial information. For one \gls{fmri} sequence $X \in \mathbb{R}^{N\times T}$, there are $N$ brain regions or \gls{rois} with $T$ time frames. To achieve a joint spatial-temporal graph, we can treat each entry of the data sequence as a node, resulting in a giant graph with $NT$ nodes and a $NT$ by $NT$ adjacency matrix. However, such a method has too large a model structure and may not be necessary for time frames where there is no long-term effect, which is often the case in trial-based cognitive tasks. Consequently, instead of applying a giant matrix, we break down a giant graph into $T$ smaller graphs, defined as joint kernels. Each kernel contains information in $N$ brain regions across $K$ time frames. 
For a better presentation of the model, we first denote a node in brain region $i$ at time frame $t$ as $V_i^t$. We then define a joint kernel graph of node $V_i^t$ as $G_i^t$. The nodes in $G_i^t$ include all brain regions that have anatomical connections to region $i$ in a $K$-frame window (i.e., $\{t-K+1, t-K+2, ..., t-1, t\}$). $X_i^t \in \mathbb{R}^{1\times 1}$ is a scalar feature associated with $V_i^t$, representing the neural activity for region $i$ at time $t$. The anatomical connections are indicated by the adjacency matrix $A$ defined by \gls{dwi}. Fig.~\ref{fig:jgatmodel}B shows the schematic of a joint kernel $G_i^t$ with $K=3$. The yellow squares are neighbors of $V_i^t$ in the joint kernel graph $G_i^t$ with anatomical connections. All the yellow nodes within the $K$-frame time window communicate with $V_i^t$. We also assume self-communication, resulting in a self-loop edge. Therefore, for node $V_i^t$, not only can it receive messages from the neighborhood (e.g., $j$ and $k$ in Fig.~\ref{fig:jgatmodel}B) in the current time, but it can also receive messages from its neighborhood from the previous time frames.

\begin{figure}[!t]
    \centering
	\includegraphics[width=1\textwidth]{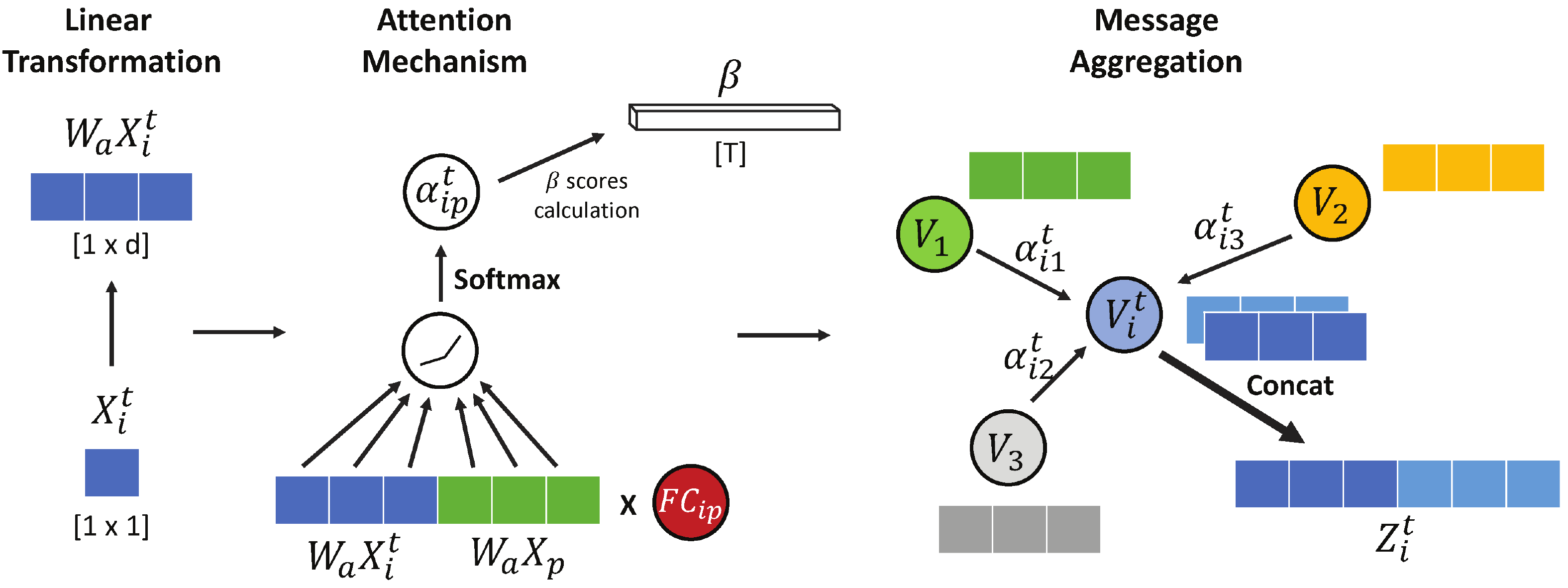}
	\caption{Overview of the JGAT layer.}
	\label{fig:jgatlayer}
\end{figure}

After defining the joint graph $G_i^t$, we now introduce the attention mechanism in the \gls{jgat} layer. The input feature for node $V_i^t$ is $X_i^t$ from the raw \gls{fmri} data. We aggregate information from its neighbors in $G_i^t$ to get a new message $\Tilde{X_i^t}$, computed as $\Tilde{X_i^t} = \sum_{p\in G_i^t}\alpha_{ip}^{t}X_{p}W_a$, where $\alpha_{ip}^{t}$ refers to the \gls{as}. It can be understood as the weight for the connection between the node $V_i^t$ and its neighbor $V_p$ in its joint domain graph $G_i^t$. Note that $p$ indexes a node in $G_i^t$. But it actually implies both the index of the brain region and the time frame in the window. Equivalently, we can consider the vectorization of nodes in the matrix form of $G_i^t$ (yellow squares in Fig.~\ref{fig:jgatmodel}B). The new index in the vector form is $p$, which corresponds to a certain region index and time index in the matrix form. $W_a \in \mathbb{R}^{1\times d}$ denotes the learnable weight matrix of linear transformation for input feature $X_{p}$. We set $d=10$ for all the experiments in this paper. The \gls{as} is defined as:
$\alpha_{ip}^{t} = Softmax(c_{ip}^{t})$, where $c_{ip}^{t}$ is the \gls{as} before the activation function. $\alpha_{ip}^t$ can be expanded with the following attention mechanism:
\begin{equation}
\alpha_{ip}^{t} = \frac{exp(f((X_{i}^{t}W_a \mathbin\Vert FC_{ip} \cdot X_{p}W_a) \cdot W_c))}{\sum_{p\in G_i^t}exp(f((X_{i}^{t}W_a \mathbin\Vert FC_{ip} \cdot X_{p}W_a) \cdot W_c))}\text{,}
\label{eq:att}
\end{equation}
where $||$ denotes vector concatenation and $W_c \in \mathbb{R}^{2d\times 1}$ denotes a query in the attention mechanism. A key for this attention mechanism is the concatenation of $X_i^t$ and $X_p$ after a linear transformation by $W_c$. The query makes the key concentrates on one value that can reflect the importance of $V_p$ with respect to node $V_i^t$. $FC_{ip}$ in Eq.~\ref{eq:att} is a Pearson correlation coefficient value that reflects the relationship between the whole time series of the region $i$ and the region corresponding to index $p$. Fig.~\ref{fig:jgatmodel}C displays the effects of \gls{fc}. For node $V_i^{t}$ in Fig.~\ref{fig:jgatmodel}C, it should have stronger connectivity with $V_j^{t-1}$ than with $V_k^{t-1}$ since the higher functional similarity defined by the Pearson correlation coefficient. By applying $FC_{ip}$ to regulate the importance of the neighborhood, we can expect the model to capture a more accurate embedding for node $V_i^t$ by considering the similarity among its neighbors. $f$ is the activation function which is $Leaky ReLU$ in our implementation. The final output embedding $Z_i^{t}$ for a single node $V_i^{t}$ is the concatenation of node representations and its aggregated message multiplied by a corresponding \gls{fs} $\beta^t$ for frame $t$, defined as $Z_i^{t} = [X_{i}^{t}W_a \mathbin\Vert \beta^t \cdot \Tilde{X_i^t}]$, where $\beta^t$ is computed from $c_{ij}^{t}$ with normalization operations:
\begin{equation}
C_i^{t} = \frac{1}{|G_i^{t}|}\sum_{p \in G_i^{t}}c_{ip}^{t},\quad \hat{\beta^{t}} = \frac{1}{N}\sum_{i=1}^{N}C_i^{t},\quad \beta^{t} = Sigmoid(\frac{\hat{\beta^{t}} - \mu(\hat{\beta^{t}})}{\sigma(\hat{\beta^{t}})}).
\label{eq:norm}
\end{equation}
$C_i^{t}$ is the mean of all neighbors of node $V_i^t$, $|G_i^{t}|$ denotes the number of neighborhood of node $V_i^t$, and $\hat{\beta^{t}}$ is the mean of $C_i^{t}$ over all brain regions in the same time frame. $\mu$ and $\sigma$ in Eq.~\ref{eq:norm} denotes the mean and standard deviation of all \gls{fs}. After normalization and the $Sigmoid$ function, we get a vector $\beta = [\beta^{1}, \beta^{2},..., \beta^{T}]^\top \in \mathbb{R}^{T\times 1}$ that reflects the importance of each frame with values ranging from 0 to 1. Fig.~\ref{fig:jgatlayer} demonstrates the overview of attention mechanism in the \gls{jgat} layer.

Finally, as mentioned above, we apply another dense layer to $Z$ to get $\tilde{Z}$ which is further fed to the readout module to build the classification loss, which is the regular categorical cross-entropy loss.

\section{Experiments and results}

\textbf{Datasets}: In this work, we use four neuroscience datasets: three of them from WU-Minn \gls{hcp} 1200 Subjects Data Release \cite{hcp} which contains preprocessed 7T fMRI data and the fourth one is from animal neural recordings \cite{animal}. For \gls{hcp} datasets, We select \gls{wm}, Social, and Emotion to do binary classification tasks. Experimental designs of each dataset typically include an experimental group and a control group to accurately measure the brain activities and regions of interest. The experimental designs usually repeat with their interested stimuli and intertwined with the control group. Hence, by following \gls{hcp} protocol, we can slice several segments of both the experimental and control groups from the fMRI time series of a single participant. The reason for selecting these three datasets is that they have uniform task blocks for both the experimental and control groups. The presence of uniform task blocks eliminates the need to crop or zero-pad to match the shape of the different cognitive stimuli (classes), thereby avoiding the risk of losing information.

For animal neural recordings, we apply our model to perform an 8-class classification task. The experimental design involved training two rhesus macaques, Chewie and Mihi, to perform a reaching task towards one of eight different directions following a cue. Data was collected from the primary motor cortex of their brains. We only present the \gls{jgat} model on neural recordings of mihi in this paper. The animal neural recordings help validate the classification performance of our \gls{jgat} model.

\textbf{Experimental setup}: For each of the \gls{hcp} datasets, the experiment includes the \gls{fmri} and \gls{dwi} data from 48 participants (subjects). After the slicing process to separate the experimental and control groups, \gls{wm} has two stimuli, 2-back \gls{wm} task and 0-back \gls{wm} task, with input shape of $M$ = 768, $N$ = 200, $T$ = 35; Social has two stimuli, Mental and Random, with input shape of $M$ = 384, $N$ = 200, $T$ = 32; Emotion has two stimuli, Fear and Neutral, with input shape of $M$ = 384, $N$ = 200, $T$ = 25. $M$ indicates the number of instances/trials. As shown in Fig.~\ref{fig:jgatmodel}A, the first stream of the model input takes advantage of \gls{dwi} data. To seize group-level features, we use a threshold to filter out some rare edges in $\tilde{A}$ and therefore preserve edges that most people have. Since each subject has one individual $\tilde{A}$, the threshold is firstly set to be 40 to represent about 80\% of participants. For animal neural recordings, there are 8 directions as labels with total inputs shape of $M$ = 215, $N$ = 172, $T$ = 30. Since neural recordings do not have data similar to DWI, we use fully connected $A$ as the adjacency matrix. 

We train and test the algorithm on Tensorflow in the Python environment using Google Colab. Furthermore, considering the relatively small datasets, the whole experiments on \gls{hcp} datasets use 8-Fold cross-validation, and the animal dataset uses 5-Fold cross-validation. In addition, the accuracy shown in the following tables takes an average of 48 samples for the HCP datasets and 30 samples for the animal dataset with \gls{sem}. The \gls{hcp} datasets are also arranged in the order of participants to prevent data leaking. The basic architecture of the \gls{jgat} model is implemented with one-head attention and one \gls{jgat} layer with the sliding window $K = 3$ and an initial learning rate of $0.001$. We also provide results of multi-head attention with two \gls{jgat} layers.

\textbf{Hyperparameter evaluation and ablation study}: The first evaluation is the size of the sliding window $K$. Tab.~\ref{tbl:k} shows the accuracy of both the training set and testing set under different model settings of $K = 1$, $3$, and $5$. In general, all \gls{hcp} datasets show better results when using $K = 3$ compared to $K = 1$. This observation supports the validity of the strategy of considering several previous time frames as neighbors. Moreover, in the Social and Emotion dataset, using $K = 3$ preserves more temporal information compared to $K = 5$. In the \gls{wm} dataset, it exhibits a longer temporal dependency, as evidenced by the similar and decent performance when using both $K = 3$ and $K = 5$.

\begin{table}[!t]
\centering
\caption{Size of joint kernel $K$.}
\label{tbl:k}
\resizebox{0.6\textwidth}{!}{%
\begin{tabular}{lllllll}
\hline
&& $K=1$ & $K=3$ &$K=5$\\ \hline
\gls{wm} & Train & $96.21\pm 0.53$  & $99.02\pm 0.30$ & $96.95\pm 0.64$  \\
         & Test  & $85.66\pm 0.69$  & $86.50\pm 0.76$ & $\mathbf{86.59\pm 0.70}$ \\ \hline

Social & Train & $99.22\pm 0.16$ & $99.81\pm 0.08$ & $99.70\pm 0.07$ \\
        & Test & $91.93\pm 0.92$ & $\mathbf{94.27\pm 0.65}$ & $93.06\pm 0.69$ \\ \hline

Emotion & Train & $98.13\pm0.29$  & $98.93\pm0.37$ & $98.71\pm0.34$  \\
        & Test & $87.76\pm0.96$  & $\mathbf{90.15\pm0.68}$ & $88.11\pm0.80$\\ \hline

\end{tabular}%
}
\end{table}

Secondly, aside from the regular attention mechanism, we also use \gls{fc} to regulate features from the neighborhood and \gls{fs} to indicate the importance of each time frame. The impact of \gls{fc} and \gls{fs} is presented in Tab.~\ref{tbl:fcfs} for all \gls{hcp} datasets with the accuracy performance of both the training set and testing set. 

\begin{table}[!t]
\centering
\caption[Ablation study on FC and FS]{Ablation study on FC and FS.}
\label{tbl:fcfs}
\resizebox{0.8\textwidth}{!}{%
\begin{tabular}{lllllll}
\hline
&& Without FC & Without FS & Without FC/FS & Full model \\ \hline
\gls{wm} & Train & $96.50\pm0.52$ & $96.18\pm0.54$  & $95.35\pm0.82$  & $99.02\pm0.30$ \\
         & Test & $86.11\pm0.71$ & $86.33\pm0.83$  & $85.94\pm0.85$  & $\mathbf{86.50\pm0.76}$ \\ \hline

Social & Train & $99.70\pm0.10$ & $99.75\pm0.07$  & $99.70\pm0.09$  & $99.81\pm0.08$ \\
        & Test & $94.14\pm0.68$ & $93.84\pm0.65$  & $93.71\pm0.64$  & $\mathbf{94.27\pm0.65}$ \\ \hline

Emotion & Train & $98.92\pm0.30$ & $99.09\pm0.17$  & $98.26\pm0.62$  & $98.93\pm0.37 $\\
        & Test & $88.32\pm0.84$ & $88.15\pm0.74$  & $88.15\pm0.98$  & $\mathbf{90.15\pm0.68}$ \\ \hline

\end{tabular}%
}
\end{table}

Thirdly, in the preliminary experimental setting, the threshold for filtering the general adjacency matrix is set to 40, resulting in a total number of approximately 5k edges in the joint kernel with $K=3$. In the following evaluation (Tab.~\ref{tbl:edges}), we demonstrate the differences in accuracy with $K=3$ under roughly 13k edges, 5k edges, 1.5k edges, and 0 edges by changing the threshold. Threshold = 24 produces roughly 13k edges, threshold = 40 produces roughly 5k edges, threshold = 48 produces roughly 1.5k edges, and \gls{mlp} model represents the model setting of 0 edges. From Tab.~\ref{tbl:edges}, we can observe that both 1.5k and 5k settings perform well, but there is no significant difference between 1.5k and 5k edges. Considering that the main objective of this work is to uncover the dynamic connectivity between brain regions and provide interpretations from an edge perspective, we use 5k edges in the final model setting in order to preserve edge information as much as possible.  

\begin{table}[!t]
\centering
\caption[Evaluation of number of edges]{Evaluation of number of edges.}
\label{tbl:edges}
\resizebox{0.8\textwidth}{!}{%
\begin{tabular}{lllllll}
\hline
&& 0 edge & 1.5k edges & 5k edges & 13k edges \\ \hline
\gls{wm} & Train & $97.93\pm 0.32$  & $99.06\pm 0.27$ & $99.02\pm 0.30$ &  $97.22\pm 0.45$\\
          & Test & $84.33\pm 0.74$ & $86.28\pm 0.75$ & $\mathbf{86.50\pm 0.76}$ &  $85.91\pm 0.63$\\ \hline

Social & Train & $99.55\pm 0.14$ & $99.63\pm 0.13$ & $99.81\pm 0.08$ & $99.48\pm 0.11$  \\
         & Test & $91.36\pm 0.59$ & $\mathbf{94.40\pm 0.62}$ & $94.27\pm 0.65$ & $93.14\pm 0.52$ \\ \hline

Emotion & Train & $99.18\pm 0.22$ &  $99.43\pm 0.17$ & $98.93\pm 0.37$  & $98.85\pm 0.22$ \\
         & Test & $88.59\pm 0.67$ & $\mathbf{90.63\pm 0.68}$ & $90.15\pm 0.68$ & $88.80\pm 0.75$ \\ \hline
\end{tabular}%
}
\end{table}

\textbf{Comparison with baseline and advanced models}: We compare our method with some basic ML models including \gls{mlp}, \gls{cnn}, \gls{gat} \cite{gat}, and some advanced methods such as GCN-LSTM, \gls{stgcn} \cite{stgcn}, \gls{stndt} \cite{stndt}, \gls{adbnn} \cite{adbnn} to evaluate the accuracy of classification results. The comparisons of the testing set are presented in Tab.~\ref{tbl:testacc}. 

\gls{mlp} employs three fully connected layers to extract features from the whole \gls{fmri} time series and all regions. 2D \gls{cnn} \cite{cnn1, cnn2} considers \gls{fmri} sequence as input and applies a filter size of $5\times 1$ along the direction of time series with two CNN layers followed by 2 dense layers. \gls{gat} is implemented as a basic spatial graph model which treats \gls{fmri} time series of each region as vector features and aggregate neighborhood messages by $A$ collected from \gls{dwi} data. GCN-LSTM model is inspired by T-GCN model which combines \gls{gcn} and \gls{rnn} models to do traffic prediction \cite{tgcn}. GCN-LSTM is treated as a dynamical graph model since \gls{gcn} and \gls{lstm} \cite{lstm} can collect information from spatial and temporal directions, though respectively. \gls{lstm} is a light version of \gls{rnn}, which has fewer parameters but can achieve similar performance as \gls{rnn}.

For the advanced models, \gls{stgcn} \cite{stgcn} is used for action recognition and prediction for the dynamics of human skeletons by combing \gls{cnn} and \gls{gcn} for several layers. \gls{stndt} \cite{stndt} employs the design of \gls{ndt} \cite{ndt} architecture and self-attention mechanism to learn spatial covariation and temporal progression from neural activity datasets. \gls{adbnn} \cite{adbnn} makes use of \gls{fc}, \gls{sc}, and \gls{gat} to consider both direct and indirect connectivity for analyzing brain activities for patients with \gls{fle} and \gls{tle}.

\begin{table}[!t]
\centering
\caption{Classification Accuracy Comparison: Baseline and Advanced \gls{ml} Models.}
\label{tbl:testacc}
\resizebox{1\textwidth}{!}{%
\begin{tabular}{lllllllll}
\hline
 & \gls{mlp} & \gls{cnn} & \gls{gat} & \gls{gcn}-\gls{lstm} & \gls{stgcn} & \gls{stndt} & \gls{adbnn} & \gls{jgat}\\ \hline
\gls{wm} & $84.33\pm0.74$ & $84.05\pm0.86$ & $83.27\pm0.55$ & $85.53\pm0.68$ & $83.62\pm0.93$ & $86.37\pm0.74$ & $64.37\pm1.07$ & $\mathbf{86.50\pm0.76}$ \\
Social & $91.36\pm0.59$ & $87.37\pm1.30$ & $88.06\pm0.85$ & $92.97\pm0.61$ & $86.63\pm0.78$ & $93.88\pm0.69$ & $72.01\pm1.20$ & $\mathbf{94.27\pm0.65}$ \\
Emotion & $88.59\pm0.67$ & $80.21\pm1.28$ & $86.28\pm0.48$ & $\mathbf{96.27\pm0.25}$ & $80.08\pm0.95$ & $89.71 \pm0.69$ & $61.63\pm0.87$ & $90.15\pm0.68$ \\
\\
mihi & $76.05\pm0.86$ & $64.34\pm2.60$ & $83.88\pm1.10$ & $78.68\pm2.00$ & $63.18\pm2.43$ & $84.11\pm1.05$ & N/A & $\mathbf{90.85\pm0.86}$ \\ \hline
\end{tabular}%
}
\end{table}

Some models share the \gls{gat} architecture such as \gls{jgat}, \gls{gat}, and \gls{stndt}. We only use one attention head with one model layer for these models in Tab.~\ref{tbl:testacc}. To evaluate the effects of multi-head attention and a deeper model structure, we also provide the comparison of three attention heads with two model layers among these models in Tab.~\ref{tbl:testacc2}.

\begin{table}[!t]
\centering
\caption{Comparison of three attention heads with two model layers.}
\label{tbl:testacc2}
\resizebox{0.5\textwidth}{!}{%
\begin{tabular}{llll}
\hline
 & \gls{gat} & \gls{stndt} & \gls{jgat}\\ \hline
\gls{wm} & $83.77\pm0.87$   & $86.07\pm0.95$  & $\mathbf{87.33\pm0.96}$  \\
Social & $91.06\pm0.90$  & $94.01\pm0.88$  & $\mathbf{95.23\pm0.96}$  \\
Emotion & $87.59\pm1.07$ & $90.02\pm1.00$ & $\mathbf{90.89\pm0.78}$ \\ 
\\ 
mihi & $82.79\pm0.86$ & $84.47\pm0.93$ & $\mathbf{87.89\pm0.48}$ \\ \hline
\end{tabular}%
}
\end{table}

\section{Interpretation of our JGAT model}

In this section, we use \gls{as} and \gls{fs} to interpret our method. \gls{as} are scores computed from learnable weights $W_a$ and $W_c$ following the attention mechanism in Eq.\ref{eq:att}. \gls{as} for each time frame can be reformed to be a sparse matrix with a dimension of $N\times 3N$, indicating the importance of each edge, given that $K$ has been fixed to be 3. \gls{fs} are parameters calculated from \gls{as} applying to each frame to regulate the relative importance of each frame. 

The interpretation process is to formulate a general representation for each cognitive stimuli by mapping \gls{as} of 200 \gls{rois} to 14 brain parcellations including both \gls{lh} and \gls{rh} according to Atlas Schaefer 7 network parcellation \cite{schaefer} (shown in the Tab.~\ref{tbl:7par}). 
The standard procedure of the first stage follows three steps:\\
$\bullet\quad$ Selecting edges in the $N\times 3N$ attention score matrix for each frame with scores above a threshold of $1.1\times (1/|G_i^t|)$. Later, taking the mean of the filtered attention matrices for all instances that belong to the same classes. We will get a group-level attention score matrix for each class.\\
$\bullet\quad$ Mapping the $N\times 3N$ group-level attention matrix to a 14 by 14 matrix following 7 brain parcellations. After the mapping process, calculating the absolute value of the difference between the two classes to represent the contrast between the two classes.\\
$\bullet\quad$ The output of the second step has dimensions of 14 by 14 by $T$. To obtain a general pattern, we calculate the average of AS across $T$ and binarize it to get an N by N binary matrix.\\
The rationale for the above steps is to incorporate the frames with both high \gls{as} and contrast between the two classes. To summarize, the first step selects frames with high scores, the second step selects frames with high contrast and the final step extracts the most representative patterns across the whole sequence with both high scores and high contrast. Most importantly, the extracted representations of all cognitive stimuli can be validated by previous literature.

\subsection{Working Memory (WM)}

\begin{figure}[!t]
    \centering
	\includegraphics[width=1\textwidth]{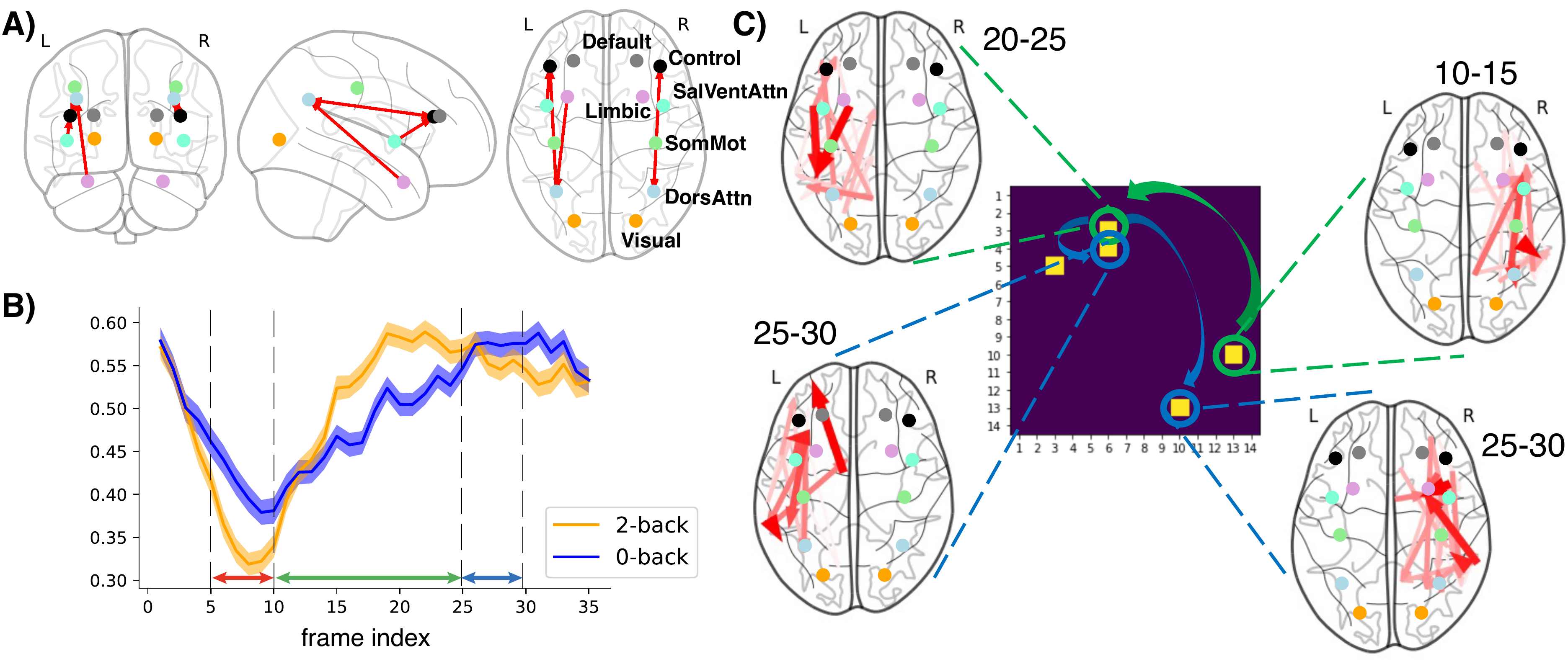}
	\caption{A) The general pattern of WM connectivity. B) \gls{fs} of WM. C) Pathway of \gls{wm} (Numbers indicate time frames).}
	\label{fig:wmgraph}
\end{figure}

In the case of WM, Fig.~\ref{fig:wmgraph}A shows the general pattern of WM which is a $14\times 14$ binary matrix. We can observe a highly activated brain network, Control(6), which constantly receives or sends messages to other brain networks including \gls{dor}(3), \gls{vent}(4), and \gls{dor}(10). Control(6) network contains a large area of \gls{pfc} that is related to \gls{wm} which has been proved by some previous literature \cite{wmval1, wmval2} and the original \gls{hcp} paper of task fMRI data \cite{hcptask}. In addition, Fig.~\ref{fig:wmgraph}B illustrates the diagram of \gls{fs} with \gls{sem}, which reveals distinguishable and significant segments that reflect the trend of time frames. We extract subgraphs from the consecutive time frames of these segments to observe the dynamical connectivity of \gls{wm} stimuli. These subgraphs form a dynamical pathway that describes the dynamical activity of \gls{wm}, as shown in Fig.~\ref{fig:wmgraph}C. The square matrix shows the binary general pattern as shown in Fig.~\ref{fig:wmgraph}A. For each highlighted yellow square, we visualize the edges in the original 200-node graph. The multiple redness levels indicate the attention scores of the edges. 

At the beginning of the \gls{wm} task, although the \gls{fs} shows that it is a distinguishable segment (red segment in Fig.~\ref{fig:wmgraph}B), it locates in the valley region, which indicates that the early stage of \gls{wm} has less influence on the downstream brain-decoding task. \gls{wm}-related neural activity is generated on the \gls{rh} indicating dominating communications between \gls{dor}(10) and \gls{vent}(11). In the middle stage, \gls{fs} shows higher values (green segment in Fig.~\ref{fig:wmgraph}B), and the subgraphs of this stage diffuse from the \gls{rh} to \gls{lh}. Most importantly, the 2-back \gls{wm} trial starts to dominate the neural activities and it reaches the peak around the 20th-25th frames, which form the most informative segment and explainable dynamical graphs. Information flows significantly from Control (6) to \gls{dor}(3) and \gls{vent}(4). In the last stage of stimuli (blue segment in Fig.~\ref{fig:wmgraph}B), the \gls{wm} task loses the retrieval of the 2-back trial and is dominated by the 0-back trial instead. The subgraphs during this period also diffuse back to the \gls{rh}, which is a similar location but in a different direction than the beginning of the middle stage. This phenomenon could indicate that the 2-back \gls{wm} is less related to the \gls{rh} since it loses the retrieval during these time intervals. According to Fig.~\ref{fig:wmgraph}B and C, there is more information flowing from the Control network to the Limbic Network in the left hemisphere between the 15th and 25th frames, indicating that it might be the pathway for 2-back \gls{wm} trials.

\subsection{Social}

\begin{figure}[!t]
    \centering
	\includegraphics[width=1\textwidth]{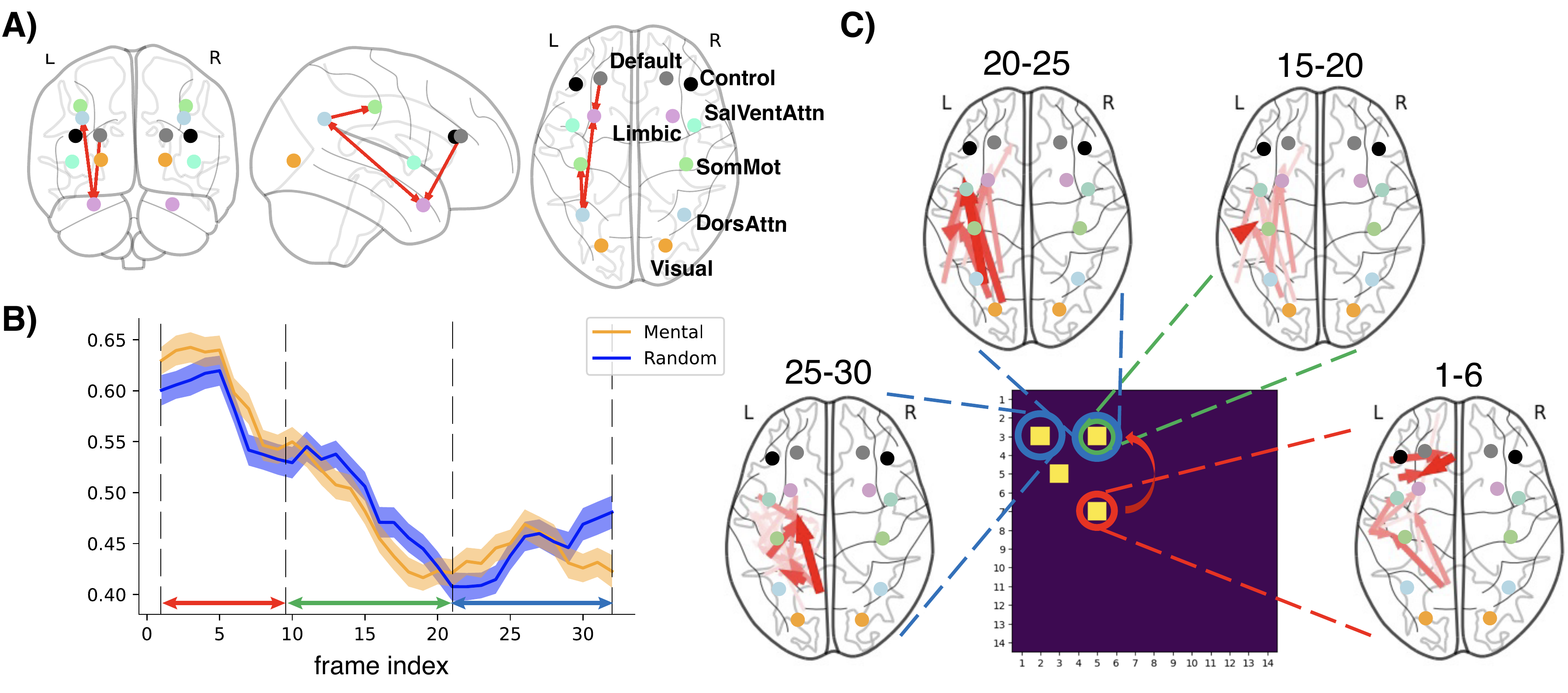}
	\caption{A) The general pattern of Social connectivity. B) \gls{fs} of Social. C) Pathway of Social (Numbers indicate time frames).}
	\label{fig:socialgraph}
\end{figure}

In the case of Social, Fig.~\ref{fig:socialgraph}A shows the general pattern of Social. From Fig.~\ref{fig:socialgraph}A, two brain parcellations, Limbic(5), Default(7), are highlighted which have a relationship with Social stimuli proven by previous literature \cite{hcptask, socialval1}. To be more specific, both Limbic(5) and Default(7) consist partially of the temporal region of the brain which is indicated to be the main target of social cognition \cite{hcptask}, especially the temporal parietal junction located at Default(7). Moreover, Fig.~\ref{fig:socialgraph}B shows the diagram of \gls{fs} with \gls{sem}, which reveals the trend of both cognitive stimuli. We also extract subgraphs from the consecutive time frames of these segments to observe the dynamical connectivity of Social. These subgraphs form a dynamical pathway that describes the dynamical activity of Social, as shown in Fig.~\ref{fig:socialgraph}C. The square matrix shows the binary general pattern as shown in Fig.~\ref{fig:socialgraph}A. For each highlighted yellow square, we visualize the edges in the original 200-node graph. The multiple redness levels indicate the attention scores of the edges. 

In the early stage of Social stimuli (red segment in Fig.~\ref{fig:socialgraph}B), due to the highest \gls{fs} and the higher retrieval of the Mental trial, it should be the most recognizable duration for the model. The informative edges concentrate in the Default network and we can also observe the Default network in this stage from Fig.~\ref{fig:socialgraph}C. There is dominating communication between Default(7) and Limbic(5). During the middle stage (green segment in Fig.~\ref{fig:socialgraph}B), \gls{fs} decreases dramatically, and the Random trial dominates as social signals decrease. This indicates that the middle stage is a less important stage for the decoding-task compared to the first stage. The edges in the subgraphs of this stage are predominantly light red, indicating low weights for these edges. Later, the Mental trial dominates for a short period again in the last stage (blue segment in Fig.~\ref{fig:socialgraph}B). However, since this short period is still around the valley of stimuli, this brief retrieval has minor importance for the decoding problem. In summary, from the early stage and the short period of the last stage in Fig.~\ref{fig:socialgraph}C, we can claim that the Default network on the \gls{lh} is critical for the Mental trial, as it communicates constantly with many other networks during the stimuli, especially Limbic. The first stage might be the best segment to describe the pathway of the Mental trial.

\subsection{Emotion}

\begin{figure}[!t]
    \centering
	\includegraphics[width=1\textwidth]{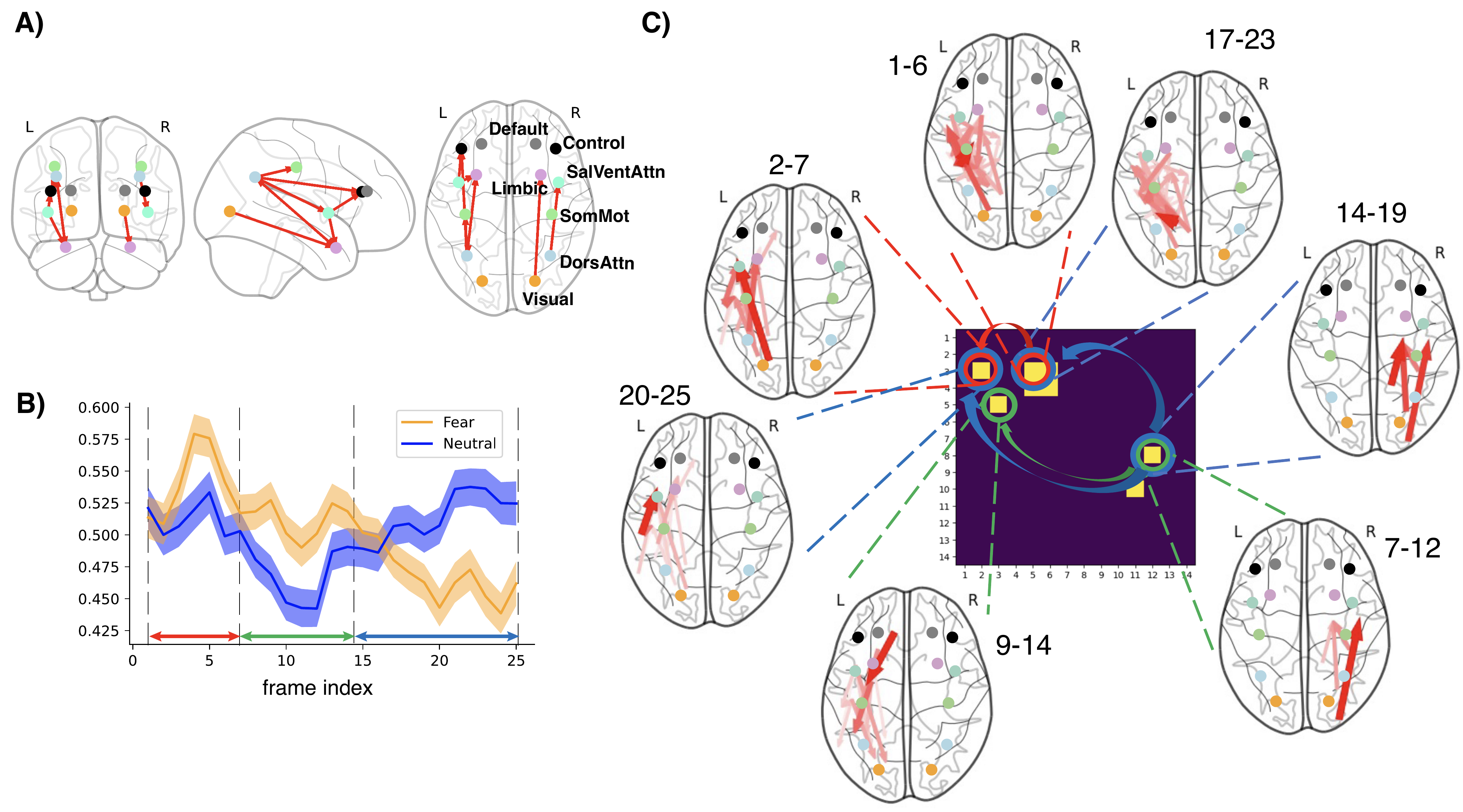}
	\caption{A) The general pattern of Emotion connectivity. B) \gls{fs} of Emotion. C) Pathway of Emotion (Numbers indicate time frames).}
	\label{fig:emotiongraph}
\end{figure}

In the case of Emotion, Fig.~\ref{fig:emotiongraph}A shows the general pattern of Emotion. From Fig.~\ref{fig:emotiongraph}A, one clear brain parcellation, Limbic(5, 12), is emphasized in both hemispheres. The Limbic contains amygdala and hippocampus which are brain regions that relate to Emotion stimuli \cite{hcptask}. Additionally, Fig.~\ref{fig:emotiongraph}B shows the diagram of \gls{fs} with \gls{sem}, which reveals the trend of both cognitive stimuli. We also extract subgraphs from the consecutive time frames of these segments to observe the dynamical connectivity of Emotion. These subgraphs form a dynamical pathway that describes the dynamical activity of Social, as shown in Fig.~\ref{fig:emotiongraph}C. The square matrix shows the binary general pattern as shown in Fig.~\ref{fig:emotiongraph}A. For each highlighted yellow square, we visualize the edges in the original 200-node graph. The multiple redness levels indicate the attention scores of the edges. 

The trend of the Emotion \gls{fs} is similar to that of the Social \gls{fs} since their highly activated segments are at the beginning of their respective stimulus tasks. During the early stage of this stimuli (red segment in Fig.~\ref{fig:emotiongraph}B), the Fear trial dominates the majority of signals and the signals clearly flow from the \gls{dor} network to the Limbic network. We can also observe the subgraph emphasize the similar brain regions during this period in Fig.~\ref{fig:emotiongraph}C. During the middle stage (green segment in Fig.~\ref{fig:emotiongraph}B), the Fear trial still dominates the response, but both Fear and Neutral trials are relatively low compared with the first stage, which leads to non-representative edges for dynamical graphs. At the end of the Emotion task (blue segment in Fig.~\ref{fig:emotiongraph}B), the main trial, Fear, drops to the valley, while the Neutral stimulus remains at the average level. The significant drops in \gls{fs} could be interpreted as fatigue towards the Fear stimulus, but it needs more experiments to confirm this interpretation. In short, the first stage of the Emotion task can best delineate the pathway of the Fear trial, which flows from the \gls{dor} network to the Limbic network. This is due to the first stage's highest \gls{fs} and the domination of the Fear trial. Furthermore, the last stage is also recognizable in the brain-decoding process given the highest contrast between the Fear/Neutral trials. Since the response of Fear trials disappears at last, it is difficult to study Fear stimuli from this segment.

\section{Conclusion}

In this paper, we propose a Joint kernel Graph Attention Network (\gls{jgat}) to integrate Functional Connectivity (\gls{fc}) and Structural Connectivity (\gls{sc}) and take temporal variations into account simultaneously. In our model, we define joint graphs for the \gls{fmri} inputs to capture edge connections across multiple time frames. Additionally, we employ Attention Scores (\gls{as}) and Frame Scores (\gls{fs}) to regulate the features of neighboring nodes within these graphs. This allows us to incorporate temporal information and optimize the representation of the data. In general, our method achieves higher accuracy when compared to both baseline \gls{ml} models and some advanced ones. Finally, by analyzing \gls{as} and \gls{fs}, we are able to provide meaningful interpretations for dynamical connectivity and construct pathways of informative segments across the time series of cognitive stimuli for the \gls{hcp} datasets. We include a discussion of limitations and future works in the appendix.



\bibliographystyle{unsrt}  
\bibliography{references}  

\newpage

\section*{Appendix}

\begin{table}[H]
\centering
\caption[7 brain networks parcellation]{Brain 7 Networks Parcellation}
\label{tbl:7par}
\resizebox{1\textwidth}{!}{%
\begin{tabular}{|c c c||c c c|}
\hline
& Left Hemisphere(\gls{lh}) &&& Right Hemisphere(\gls{rh}) &\\ \hline
Number & Network  & Nodes range  & Number & Network  & Nodes range  \\ \hline
1 & Visual & 1 – 14 & 8 & Visual & 101 - 115 \\
2 & SomMot & 15 – 30 & 9 & SomMot & 116 - 134 \\
3 & \gls{dor} & 31 – 43 & 10 & \gls{dor} & 135 - 147 \\
4 & \gls{vent} & 44 – 54 & 11 & \gls{vent} & 148 - 158 \\
5 & Limbic & 55 – 60 & 12 & Limbic & 159 - 164 \\
6 & Control & 61 – 73 & 13 & Control & 165 - 181 \\
7 & Default & 74 – 100 & 14 & Default & 182 - 200 \\ \hline

\end{tabular}%
}
\end{table}

\subsection*{Effect of FC and FS}

\begin{figure}[H]
     \centering
     \begin{subfigure}[b]{0.8\textwidth}
         \centering
         \includegraphics[width=\textwidth]{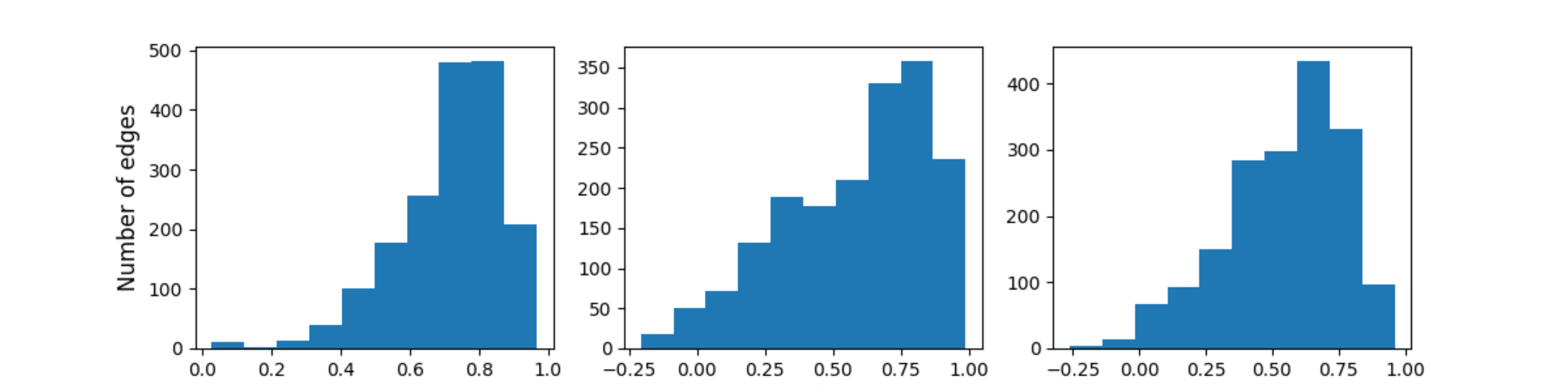}
         \caption{Functional Connectivity (FC)}
         \label{fig:fc_hist}
     \end{subfigure}
     \hfill
     \begin{subfigure}[b]{0.8\textwidth}
         \centering
         \includegraphics[width=\textwidth]{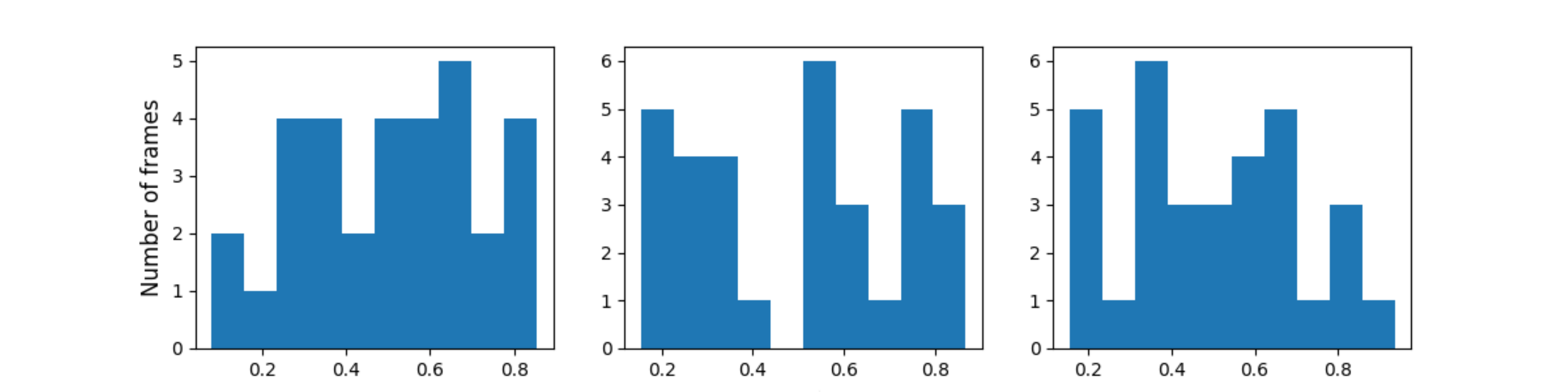}
         \caption{Frame Scores (FS)}
         \label{fig:fs_hist}
     \end{subfigure}
        \caption{Histogram of \gls{fc} and \gls{fs}}
        \label{fig:his_fcfs}
\end{figure}

From the modeling perspective, we involve \gls{fc} and \gls{fs} to regulate the dynamical features. We also do the ablation study on these two scores in Tab.~\ref{tbl:fcfs}. By observing the results, it seems that for the same datasets such as \gls{wm} and Social, there does not exist a significant difference by applying these two scores. However, we can move a step back to visualize the histogram of two scores first in Fig.~\ref{fig:his_fcfs}. The plots show the histograms of \gls{fc} and \gls{fs} from some random Social instances. These plots indicate that these scores do have certain distributions instead of barely uniform values. After employing two scores with certain distributions, the results do not drop and some of them even improve. Furthermore, the \gls{fs} is useful in the interpretation part as it helps identify numerous meaningful edges and brain parcellations. 

\subsection*{Limitations and Future Works}

There are several limitations related to the \gls{jgat} model. First of all, we can only provide group-level interpretation for this model at this stage. The individual explanation is a more complex topic, but it is certainly worth exploring and delving into due to its significance and intriguing nature. One of the reasons for the presence of noise in our interpretation is the significant influence of individual differences among all instances when averaging our results. Therefore, providing individual explanations has the potential to yield clearer dynamical graphs and results by accounting for the specific characteristics and variations present in each individual's data. Some thoughts on individual interpretation could involve utilizing individual adjacency matrices instead of a general one or adding additional loss to the control group or individual level. Secondly, running time is an issue for this model. Due to the large number of edges in our method, which is three times more than other models, and the inclusion of multiple scores to enhance the model, our structure appears to be the most complex when compared to other models used in the comparison. In the future, it would be better to optimize the algorithm and codes. Lastly, for datasets that lack an adjacency matrix or similar information, such as the mihi dataset, the only option is to use a fully connected adjacency matrix as input. In such cases, the model does not have access to specific connectivity information and treats all brain regions as fully connected. Indeed, using a fully connected adjacency matrix can pose challenges when trying to generate a meaningful interpretation using the method we employed. In essence, datasets of this type may not be suitable for our current method, or it may be necessary to develop an alternative interpretation approach specifically tailored for fully connected adjacency matrices.


\end{document}